\documentclass[clsnofoot,10pt,twocolumn,letterpaper]{IEEEtran}
\usepackage{graphicx}
\usepackage{amsmath}
\usepackage{amssymb}
\usepackage{algorithm}
\usepackage{algorithmic}
\usepackage{array}
\usepackage{multicol}
\usepackage{multirow}
\usepackage{soul}

\begin{document}
\title{Deep Learning based Joint Precoder Design and Antenna Selection for Partially Connected Hybrid Massive MIMO Systems}
\author{Salman Khalid, Waqas~bin~Abbas, Farhan Khalid, \textit{Member, IEEE}
	\thanks{
		Salman Khalid (corresponding author, email: salmankhalid16@yahoo.com), W.~bin~Abbas and Farhan Khalid are with the National University of Computer and Emerging Sciences (NUCES), Islamabad, Pakistan.}
	}
\maketitle
\begin{abstract}
Efficient resource allocation with hybrid precoder design is essential for massive MIMO systems operating in millimeter wave (mmW) domain. Owing to a higher energy efficiency and a lower complexity of a partially connected hybrid architecture, in this letter, we propose a joint deep convolutional neural network (CNN) based scheme for precoder design and antenna selection of a partially connected massive MIMO hybrid system. Precoder design and antenna selection is formulated as a regression and classification problem, respectively, for CNN. The channel data is fed to the first CNN network which outputs a subset of selected antennas having the optimal spectral efficiency. This subset is again fed to the second CNN to obtain the block diagonal precoder for a partially connected architecture. Simulation results verifies the superiority of CNN based approach over conventional iterative and alternating minimization (alt-min) algorithms. Moreover, the proposed scheme is computationally efficient and is not very sensitive to channel irregularities.  
\end{abstract}
\section{Introduction}
Researchers are exploring the millimeter (mmW) and terahertz domain to meet the ever increasing demand of wider bandwidth and higher data rates \cite{ref_01}. The propagation environment at such higher frequencies is suffered by severe path loss, scattering and penetration losses. Massive MIMO architecture with precoding/beamforming gain is utilized to compensate for the propagation losses \cite{ref_02}. Researchers are tilted towards the hybrid (analog coupled with digital) beamforming architecture since it provides the gains of digital processing with lower power consumption \cite{ref_03}\cite{ref_04}. Existing literature proposes many techniques for both fully connected (where each RF chain is connected to every antenna) and partially connected (where each RF chain is connected to a subset of antennas) hybrid architectures \cite{ref_03}-\cite{ref_07}. Authors in \cite{ref_05} have utilized the orthogonal matching pursuit (OMP), a greedy algorithm, for the computation of analog and digital precoders for a fully connected hybrid architecture by utilizing the array responses of the transmitter and the receiver. \cite{ref_06} proposes the iterative successive interference cancellation (SIC) algorithm for computation of hybrid precoder for an energy efficient partially connected hybrid architecture. Authors in \cite{ref_07} have proposed the manifold optimization (MO) and the phase extraction (PE) based alternating minimization (alt-min) techniques to compute the hybrid precoder for a fully connected architecture and the semi definite relaxation (SDR) based technique for a partially connected architecture. Alt-min approaches explores the linkage between optimal and hybrid precoders to estimate the digital and analog precoder. The application of evolutionary algorithms for evaluation of hybrid precoders is demonstrated in \cite{ref_08}\cite{ref_09}. For Massive MIMO systems, the resource allocation in terms of active antennas selection is critical to ensure high energy efficiency. The spectral efficiency gain becomes constant beyond a certain number of antennas, hence to optimize the hardware, antennas experiencing good channel conditions should be selected. For antenna selection problem, authors in \cite{ref_10} and \cite{ref_11} have applied an iterative evolutionary and estimation of distribution based algorithm. The above mentioned iterative, greedy and alt-min techniques regarding antenna selection and precoding have drawbacks in terms of computational time and achieving optimal solution in terms of spectral efficiency.

All above works on antenna selection and precoding gives the sub optimal solution with considerable computational complexity despite applying various optimization strategies and selection criteria. Recently, the machine learning based convolutional neural network (CNN) methods have gained interest of researchers to solve the optimization problems related to wireless communication. Well trained CNNs have ability to deduce features from a given set of observations with high efficiency and at a very low complexity as compared to conventional techniques. CNNs find its applications for solving problems such as channel estimation \cite{ref_12}, interference coordination, beam management \cite{ref_13} and analog beam selection \cite{ref_14}. Very recently precoder design problem is also formulated and solved using CNN \cite{ref_15}-\cite{ref_19}. However, all the research for precoder and combiner design is limited to a fully connected hybrid architecture and do not consider the partially connected hybrid architecture which has proven ability of being energy efficient and bears reduced complexity \cite{ref_06}.     

Keeping in view the low latency, low power consumption and high energy efficiency requirements of future 5G and B5G communication networks, in this letter, we propose a CNN based joint antenna selection and precoder design for a partially connected hybrid structure. Two separate CNNs i.e., classification and regression are trained for an antenna selection and a precoding problem, respectively. Channel realizations added with noise are used to train the networks and estimate the optimum antenna subset and precoder weights. The input of the channel matrix is fed to the first stage CNN deployed for an antenna selection. The reduced subset is further fed to second stage CNN which outputs the optimum analog precoder. The training of both CNNs is performed offline, hence, all the computational overhead for the data generation and training is not present during online prediction, where the hybrid precoders prediction and antennas classification is done by only feeding the channel matrix to the network.
\section{Hybrid Massive MIMO System Model}
For joint estimation of precoder weights and antenna selection, in this letter, we have considered a hybrid architecture in a partially connected configuration i.e, each RF chain energizes only a subset of antennas $M$ = $N_{T}$/$N^{RF}_{T}$. We have considered the base station (BS) equipped with $N_{T}$ transmit antennas and $N^{RF}_{T}$ RF chains to transmit $N_{S}$ data streams. The user is considered to be equipped with $N_{R}$ receive antennas where the selection is performed to determine $N_{r}$ best antennas. The analog and baseband precoder at the transmiter are represented as $\mathbf{F}_{RF} \in C^{N_{T} \times N^{RF}_{T}}$ and $\mathbf{F}_{BB} \in C^{N^{RF}_{T} \times N_{S}}$ respectively. Therefore, the signal transmitted through BS is represented as $\mathbf{x} = \mathbf{F}_{RF} \mathbf{F}_{BB} \mathbf{s}$, where $\mathbf{s}$ is the $N_{S} \times 1$ transmitted symbol vector. The analog precoder $\mathbf{F}_{RF}$ is a block diagonal matrix realized by phase shifters having equal magnitude with variable phases and satisfying the power constraint $\left\| \mathbf{F}_{RF}\mathbf{F}_{BB} \right\|_{F} \leq N^{RF}_{T}$. The received signal $\mathbf{y}$ at the user having $N_{R}$ antennas is expressed as

\begin{equation}\label{eq:op}
\mathbf{y} = \sqrt{P_{av}} \mathbf{H}\mathbf{F}_{RF}\mathbf{F}_{BB}\mathbf{s} + \mathbf{n}
\end{equation}

The average received power is $P_{av}$, $\mathbf{n}$ ($\mathcal{CN}(0,\sigma^2)$) is i.i.d complex Gaussian noise. $\mathbf{H}$ denotes the $N_{R}$ $\times$ $N_{T}$ full array channel between the transmitter and receiver. The clustered geometric Saleh-Valenzuela model \cite{ref_03} representing the low rank mmW channel is used in this letter.

\begin{equation}\label{eq:chan}
\mathbf{H} = \sqrt{\left(\dfrac{N_{T} N_{R}}{\epsilon K}\right)} \sum_{k=0}^{K}\eta_{l}\textbf{a}_{R}(\mu_{k})\textbf{a}^{H}_{T}(\theta_{k})
\end{equation}

where $K$ is the number of paths, $\epsilon$ is the pathloss, the path gain linked with the $k^{\text{th}}$ path is $\eta_{k}$, the corresponding spatial signatures of the receiver and the transmitter are $\textbf{a}_{R}$ and $\textbf{a}_{T}$, respectively, and $\mu_{k}$ and $\theta_{k}$ are the angle of arrival (AoA) and the angle of departure (AoD) of the $k^{\text{th}}$ path, respectively. Finally, the full array (without any antenna selection) spectral efficiency is defined as

\begin{equation}
	R = \log (\mathbf{I}_{N_{R}} + \frac{\rho}{N_{s}} \mathbf{H}\mathbf{F}_{RF}\mathbf{F}_{BB}\mathbf{F}^{H}_{BB}\mathbf{F}^{H}_{RF}\mathbf{H}^{H})
\end{equation}

\section{Joint Antenna Selection and Hybrid Precoder}
Our first goal is to determine a subset of $N_{r}$ best antennas out of total available $N_{R}$ receive antennas. After the antenna selection, the RF and baseband precoders are determined using reduced dimensions. The joint solution of antenna selection and precoding have to satisfy following condition

\begin{equation}
	\underset{\mathbf{Q},\mathbf{F}_{RF},\mathbf{F}_{BB}}{\max}\ \log (\mathbf{I}_{N_{r}} + \frac{\rho}{N_{s}} \mathbf{H}_{sel}\mathbf{F}_{RF}\mathbf{F}_{BB}\mathbf{F}^{H}_{BB}\mathbf{F}^{H}_{RF}\mathbf{H}^{H}_{sel})
\end{equation}

Here $\mathbf{H}_{sel}$ = $\mathbf{Q}\mathbf{H}$ is a reduced dimension ($N_{r}$ $\times$ $N_{T}$) channel matrix obtained by performing antenna selection. $\mathbf{Q}$ is a ($N_{r}$ $\times$ $N_{R}$) selection matrix with entries either $0$ or $1$ representing the antenna index.

\begin{figure*}[tbph]
\centering
\includegraphics[height=1.4in,width=7.2in]{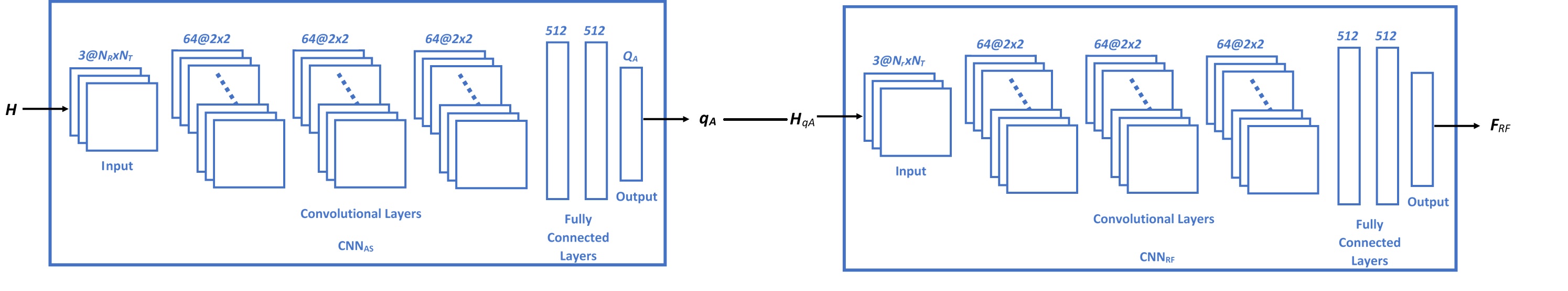}
\caption{Joint CNN Architecture for Antenna Selection and Precoding}
\label{fig:FC}
\end{figure*}

\subsection{Antenna Selection}
For antenna selection, picking $N_{r}$ antennas out of $N_{R}$ yields $\mathbf{Q}_{A}$ = $N_{R} \choose N_{r}$ possible combinations. Hence, selecting a subset of antennas becomes a CNN classification problem with $\mathbf{Q}_{A}$ classes. Let $q_{A}$th antenna subset configuration with $q_{A}$ $\in$ $\mathcal{Q}_{A} = \{1,....,Q_{A}\}$ is selected, than the received signal vector with $q_{A}$th selected subset with $\mathbf{H}_{q_{A}}$ ($N_{r}$ $\times$ $N_{T}$) being corresponding channel matrix is expressed as

\begin{equation}\label{eq:op}
\mathbf{y}_{q_{A}} = \sqrt{P_{av}} \mathbf{H}_{q_{A}}\mathbf{F}_{RF}\mathbf{F}_{BB}\mathbf{s} + \mathbf{n}_{q_{A}}
\end{equation}

Similarly, the spectral efficiency with $q_{A}$th selected subset is expressed as

\begin{equation}
	R = \log (\mathbf{I}_{N_{r}} + \frac{\rho}{N_{s}} \mathbf{H}_{q_{A}}\mathbf{F}_{RF}\mathbf{F}_{BB}\mathbf{F}^{H}_{BB}\mathbf{F}^{H}_{RF}\mathbf{H}^{H}_{q_{A}})
\end{equation}

Note that $R$ is dependent on $q_{A}$ through $\mathbf{H}_{q_{A}}$. By maximizing $R$ for all combinations of antenna selection configurations, the best antenna subset is expressed as

\begin{equation}
	 q_{A} = \underset{q_{A} \in \mathcal{Q}_{A}}{\arg\max}\ R(q_{A})
\end{equation}

\subsection{Partially Connected Hybrid Precoder Design}
Let $\mathbf{H}_{q_{A}}$ is the reduced dimensions channel matrix obtained after performing antenna selection than the hybrid precoder problem is defined as

\begin{equation}
	\underset{\mathbf{F}_{RF},\mathbf{F}_{BB}}{\max}\ \log (\mathbf{I}_{N_{R}} + \frac{\rho}{N_{s}} \mathbf{H}_{q_{A}}\mathbf{F}_{RF}\mathbf{F}_{BB}\mathbf{F}^{H}_{BB}\mathbf{F}^{H}_{RF}\mathbf{H}^{H}_{q_{A}})
\end{equation}

We have considered a partially connected hybrid structure where each RF chain is connected to $M$ = $N_{T}$/$N^{RF}_{T}$ number of antennas. This implies that the structure of the RF precoder $\mathbf{F}_{RF}$ must be a block diagonal with $\mathbf{f}_{RF_{i}}$ being the precoding vector for the \textit{i}th RF chain only having M non zero elements and is expressed as

		\begin{equation}
		\mathbf{F}_{RF} = 
		\begin{pmatrix}
		\mathbf{f}_{RF_{1}} & \mathbf{0} & \cdots & \mathbf{0} \\
		\mathbf{0} & \mathbf{f}_{RF_{2}} & \cdots & \mathbf{0} \\
		\vdots  & \vdots  & \ddots & \vdots  \\
		\mathbf{0} & \mathbf{0} & \cdots & \mathbf{f}_{RF_{N_{rf}}}
		\end{pmatrix}
		\end{equation}  

Hybrid precoder has to meet two constraints; C1: All non-zero elements of $\mathbf{F}_{RF}$ must have the same amplitude and, C2: Meet the total power constraint $\left\| \mathbf{F}_{RF}\mathbf{F}_{BB} \right\|_{F} \leq N^{RF}_{T}$.
For the case of hybrid precoder, based on the proof \cite{ref_05}, the Euclidean distance between the optimal unconstrained precoder and the hybrid precoder should be minimized. In other words, the hybrid precoder design problem is rewritten as

\begin{equation}
	\underset{\mathbf{F}_{RF},\mathbf{F}_{BB}}{\arg \min}\ \left\| \mathbf{F}_{opt} - \mathbf{F}_{RF}\mathbf{F}_{BB} \right\|^{2}_{F}
\end{equation} 

The optimal solution for the above mentioned optimization problem can be obtained using the singular value decomposition performed on the channel matrix, which can be used to generate the labels for output layer of regression CNN network. Even with the memory-friendly approaches, it is computationally complex to enumerate over all possible antenna selection subsets in real time. Also the optimal design of hybrid precoders requires iterations and extensive computations. In order to address this issue, we have formulated a deep CNN based solution where the networks are trained offline and perform computations for antenna subset configuration and optimal precoder. Afterwards, the trained network can simply be deployed as a classification and regression network to select antennas and estimate hybrid precoders.

\section{CNN Training and Dataset Generation}
Our proposed deep neural network consists of two separate CNNs (Fig. \ref{fig:FC}) to perform antenna selection and precoding. The input to the first CNN$_{AS}$ is the full dimension channel matrix which selects the best antenna subset $q_{A}$. The second CNN$_{RF}$ accepts the reduced dimensions channel matrix with only selected rows corresponding to the selected antennas and estimates the RF precoder at its output. For both architectures, the training data is generated using channel realizations which are further assigned with corresponding output class/label for antenna selection and hyrid precoding.

Let the input data $\mathbf{X}$ be a $N_{R}$$\times$$N_{T}$$\times$3 with \textit{c} = 3 channels. The first channel of input is the absolute value of imperfect channel matrix $\tilde{\mathbf{H}}$ whereas the real and imaginary part of channel matrix are stored in second and third channels, respectively. For data generation, $N$ different channel matrix realizations are generated. Afterwards for each realization, $L$ noisy channel matrices are created with synthetic noise which is added element wise. Hence, the total size of training input data becomes $N_{R}$$\times$$N_{T}$$\times$3$\times$$NL$. In order to obtain the output labels, for antenna selection CNN$_{RF}$ the best antenna subset is selected and afterwards for second CNN$_{RF}$, $\mathbf{F}_{RF}$ is obtained by performing SVD operation on reduced dimension channel realizations. Hence, the input/output pairs are established. The training process of both CNNs is identical but with different input dimensions.

The input sizes of CNN$_{AS}$ and CNN$_{RF}$ is $N_{R}$$\times$$N_{T}$$\times$3 and $N_{r}$$\times$$N_{T}$$\times$3, respectively. Each CNN is composed of 14 layers. Input layer being the first layer of corresponding input data size. The convolutional layers are second, fourth and sixth with 64 filters of dimensions 2$\times$2. Eighth and eleventh layers are fully connected layers with 512 nodes. The tenth and thirteenth layers are dropout layers with 50\% probability. The RELU activation function is utilized. Finally, the output layer of CNN$_{AS}$ is a classification layer with softmax function to output the antenna subset class which gives maximum spectral efficiency and output layer of CNN$_{RF}$ is a regression layer of dimensions $N_{T}$ $\times$ 1 representing the non zero elements of $\mathbf{F}_{RF}$. After estimating the non zeros elements of $F_{RF}$, the block diagonal structure is obtained by appending zeros at appropriate locations. CNN$_{RF}$ is used to predict the $F_{RF}$ and the $F_{BB}$ is obtained using equivalent channel approach.

\section{Numerical Simulations and Results}
In this section, we evaluate the performance of our proposed approach. The performance of CNN based hybrid precoder is evaluated against state of the art SDR alt-min and SIC algorithms. Uniform planner array with $N_{T}$ = 36 or 144 and $N_{R}$ = 16 for the transmitter and receiver respectively, are generated. For antenna selection, the $N_{r}$ is kept as 8. The RF chains at the transmitter $N^{RF}_{T}$ and at the receiver $N^{RF}_{R}$ are kept as 4. For CNNs, the training data is generated for \textit{N} = \textit{L} = 100 realizations. The proposed network is trained using MATLAB as a simulation environment. SGD algorithm is used for network parameters with learning rate of 0.005 and mini-batch size 500 with 200 epochs. The cross entropy function is used as the loss function. During the training phase, 30\% and 70\% of all data is divided into validation and training datasets, respectively. Finally the validation data is used to verify the performance of the proposed architecture in the simulations for 100 Monte Carlo trials.

\begin{figure}[tbph]
\centering
\includegraphics[height=3.2in,width=3.6in]{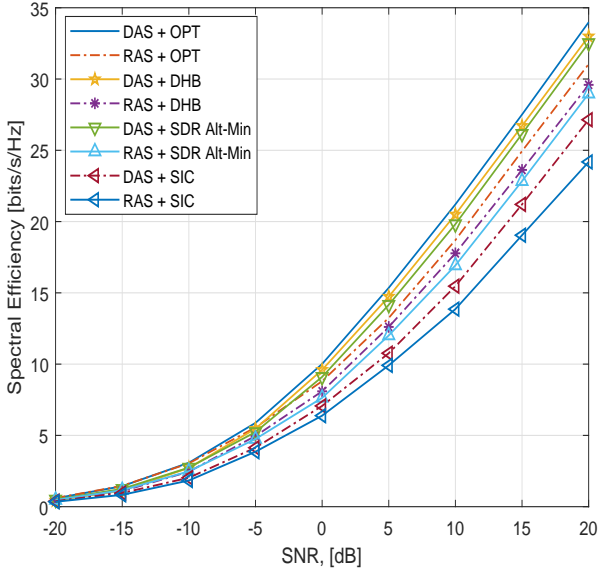}
\caption{Spectral Efficiency with $N_{T}$=36, $N_{R}$=16, $N_{r}$=8, $N^{T}_{RF}$=4}
\label{fig:res1}
\end{figure}

\begin{figure}[tbph]
\centering
\includegraphics[height=3.2in,width=3.6in]{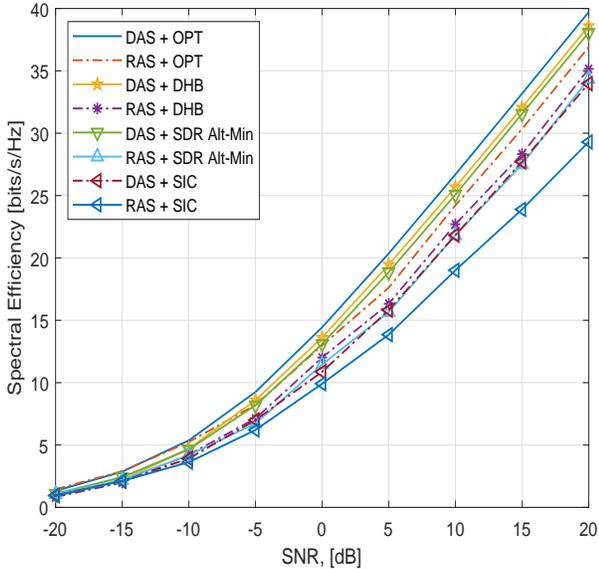}
\caption{Spectral Efficiency with $N_{T}$=144, $N_{R}$=16, $N_{r}$=8, $N^{T}_{RF}$=4}
\label{fig:res2}
\end{figure}

Fig. \ref{fig:res1} and Fig. \ref{fig:res2} with $N_{T}$ as 36 and 144 respectively, shows the spectral efficiency for different algorithms with deep antenna selection (DAS). The $N^{T}_{RF}$ and $N_{S}$ are considered equal and set to 4, the $N_{R}$ is set as 16 and DAS CNN network selects $N_{r}$ = 8 antennas. After performing DAS, the hybrid precoders are determined. CNN$_{RF}$ is used to determine the $F_{RF}$ whereas the $F_{BB}$ is obtained using equivalent channel approach. The CNN based hybrid precoder is outperforming the SDR alt-min and SIC algorithms. The CNN based hybrid precoder is efficiently predicting the RF precoder which contributes towards maximization of spectral efficiency. To evaluate the antenna selection technique, the performance of CNN based antenna selection is compared with random antenna selection (RAS) applied with precoding. It is evident that spectral efficiency using RAS algorithm is trailing behind DAS algorithm irrespective of precoding technique.

The computation time of CNN based precoder, SDR alt-min and SIC algorithms are also computed for $N_{T}$ = 144. The CNN based precoder only required 0.01s, SDR alt-min requires 1.6s and SIC based precoder requires 0.02s for computations. The CNN based hybrid precoder is outperforming all algorithms in terms of spectral efficiency. Hence, both the computational efficiency and spectral efficiency of the proposed CNN based hybrid precoder is established.    
\section{Conclusions}
This letter presents the CNN based solution for joint hybrid precoder design of a partially connected mmW massive MIMO system with antenna selection. The proposed novel technique outperforms the existing algorithms for a partially connected hybrid precoder design both in terms of spectral efficiency and computational complexity. Moreover the antenna selection enables efficient resource allocation for systems with large antenna arrays.

\end{document}